\documentclass{appolb}
\usepackage{epsf,amsmath,amssymb}
\usepackage[nosort]{cite}
\newcommand{\lmp}{large momentum procedure}
\newcommand{\hmp}{hard mass procedure}
\newcommand{\la}{\langle}
\newcommand{\ra}{\rangle}
\newcommand{\logqmms}{l_{qm}}

\newcommand{\logqmu}{l_{q\mu}}

\newcommand{\logmum}{l_{\mu m}}

\newcommand{\opo}{{\cal O}}
\newcommand{\bare}{{\rm B}}
\newcommand{\fprop}[3]{{#1\over #2^2 + #3^2}}
\newcommand{\insdia}[4]{\raisebox{#1}{\epsfxsize=#2 \epsffile[#3]{#4}}}
\newcommand{\nnb}{\nonumber}
\newcommand{\dd}{{\rm d}}

\newcommand{\oddo}[2]{#2{\dd#1\over \dd #2}}
\newcommand{\doverd}[1]{{\partial\over \partial #1}}

\newcommand{\order}[1]{{\cal O}(#1)}
\newcommand{\bld}[1]{\boldmath{$#1$}}

\renewcommand{\Im}{{\rm Im}}
\newcommand{\msbar}{{\overline {\rm MS}}}
\newlength{\fsize}
\begin{document}

\pagestyle{plain}
\title{
Asymptotic Expansions --- Methods and Applications
\thanks{Presented at the XXIII~School of Theoretical Physics, Ustro\'n99.}
}
\author{
\begin{flushright}
\vspace{-10em}
  \rm BNL-HET-99/36\\
  \rm TTP99-46\\
  \rm October 1999\\
  \rm hep-ph/9910496\\[7em]
\end{flushright}
Robert Harlander
\address{HET, Physics Department, Brookhaven National Laboratory, Upton,
  NY 11973}}
\maketitle


\begin{abstract}
Different viewpoints on the asymptotic expansion of Feynman diagrams
are reviewed. The relations between the field theoretic and diagrammatic
approaches are sketched. The focus is on problems with large masses
or large external momenta. Several recent applications also for other
limiting cases are touched upon. Finally, the pros and cons of the different
approaches are briefly discussed.
\end{abstract}


\section{Introduction}\label{sec::intro}
Feynman diagrams are the most important theoretical tool for particle
physicists. They are an efficient link between theory and experiment.
However, their translation into actual numerical predictions is often
very tedious if not impossible. Huge efforts have been devoted to
their evaluation, and several powerful methods have been developed to
systemize their treatment. The more complex a Feynman diagram is, the
more important is it to find approximation procedures that allow to solve
the problem with finite but reasonable accuracy. In this paper we will
describe methods that have been developed over the recent years in order
to systematically expand Feynman diagrams in their external parameters.


\section{Status of multi-loop calculations}\label{sec::status}
The complexity of a Feynman diagram with a certain number of loops
mainly depends on its number of scales (i.e., masses and external
momenta). In the one-loop case, the problem can be considered as solved.
Any tensor integral can be reduced to integrals of unit numerator which
have been studied extensively.  Nowadays there are powerful software
tools, on the one hand concerned with the tensor reduction, on the other
hand with the numerical or analytical evaluation of the integrals (see,
e.g., \cite{hahn}).

At two-loop level the solution is not as general as in the one-loop case.
However, the important class of two-point functions is well under control,
and the development for three- and four-point functions is under continuous
progress (see \cite{HS:review} for a list of references).

Therefore, two-loop calculations in theories like the electro-weak
standard model and even in supersymmetric models whose particle spectra
give rise to Feynman diagrams with several different scales have become
feasible~(e.g.~\cite{EW2l}).

Calculations at three-loop level mostly reside on two different classes of
analytically solvable Feynman integrals. These two classes are
\begin{itemize}
\item massless propagator-type diagrams where all internal lines are
  massless and only one external momentum is different from
  zero. Schematically:
  \begin{eqnarray}
    I(n_1,\ldots,n_\alpha) = \int\dd^Dk_1\dd^Dk_2\dd^Dk_3{P(q,k_1,k_2,k_3)\over
      (p_1^2)^{n_1}\cdots (p_\alpha^2)^{n_\alpha}}\,,
    \label{eq::prop}
  \end{eqnarray}
  where $\alpha$ is the number of propagators and the $p_i$ are linear
  combinations of the $k_j$ and the external momentum $q$.
\item massive tadpole diagrams not carrying any external momenta and
  internal lines being either massless or carrying a common mass $m$.
  Schematically:
  \begin{eqnarray}
    J(n_1,\ldots,n_\alpha) = \int\dd^Dk_1\dd^Dk_2\dd^Dk_3{P(k_1,k_2,k_3)\over
      (m_1^2 + p_1^2)^{n_1}\cdots (m_\alpha^2 + p_\alpha^2)^{n_\alpha}}\,,
    \label{eq::tad}
  \end{eqnarray}
  where the $m_i$ are either equal to zero or $m$, and the $p_i$ are
  linear combinations of the loop momenta $k_j$.
\end{itemize}
$P(\ldots)$ is a polynomial of products of its arguments.
The method how to solve such integrals is called the
integration-by-parts algorithm~\cite{IP}. It is based on identities
derived from the fact that the $D$-dimensional integral over a total
derivative is equal to zero:
\begin{equation}
\int \dd^D p\,{\partial\over \partial p_\mu} f(p,\ldots) = 0\,.
\end{equation}
These identities can be arranged in such a way that they yield
recurrence relations that allow to reduce some of the ``indices''
$n_1,\ldots,n_\alpha$ in (\ref{eq::prop}), (\ref{eq::tad}) to zero. At
three-loop level, these relations have been derived in \cite{IP} for
massless propagators and in \cite{Bro92,avdeev} for massive tadpole integrals.
Their application to a general three-loop diagram may generate huge
intermediate expressions that easily exceed several hundreds of megabytes
on a computer.  This is why one needs to implement the relations to
powerful computer algebra systems like {\tt FORM} or {\tt REDUCE}.  Two
such implementations are {\tt MINCER}~\cite{mincer}, concerned with the
massless propagator diagrams and {\tt MATAD}~\cite{Ste:diss}, dealing
with the massive tadpoles (see \cite{HS:review} for a review on
automatic computation of Feynman diagrams).

\begin{figure}
  \begin{center}
    \begin{tabular}{lll}
  \leavevmode
  (a) & (b) & (c)\\
\insdia{-1.3em}{8em}{140 265 440 470}{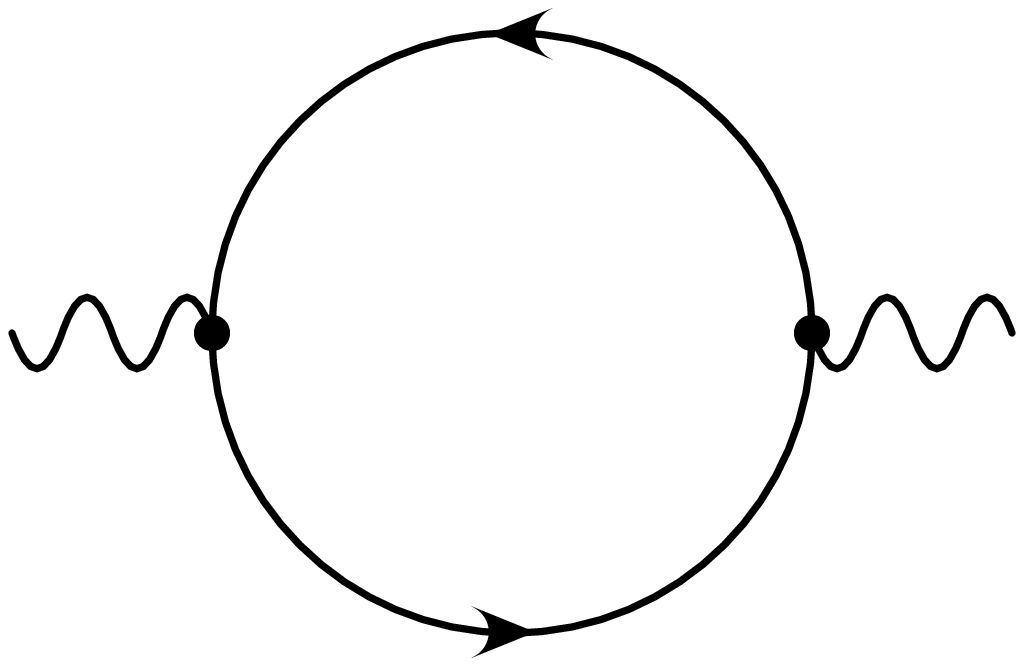} &
\insdia{-1.3em}{8em}{140 265 440 470}{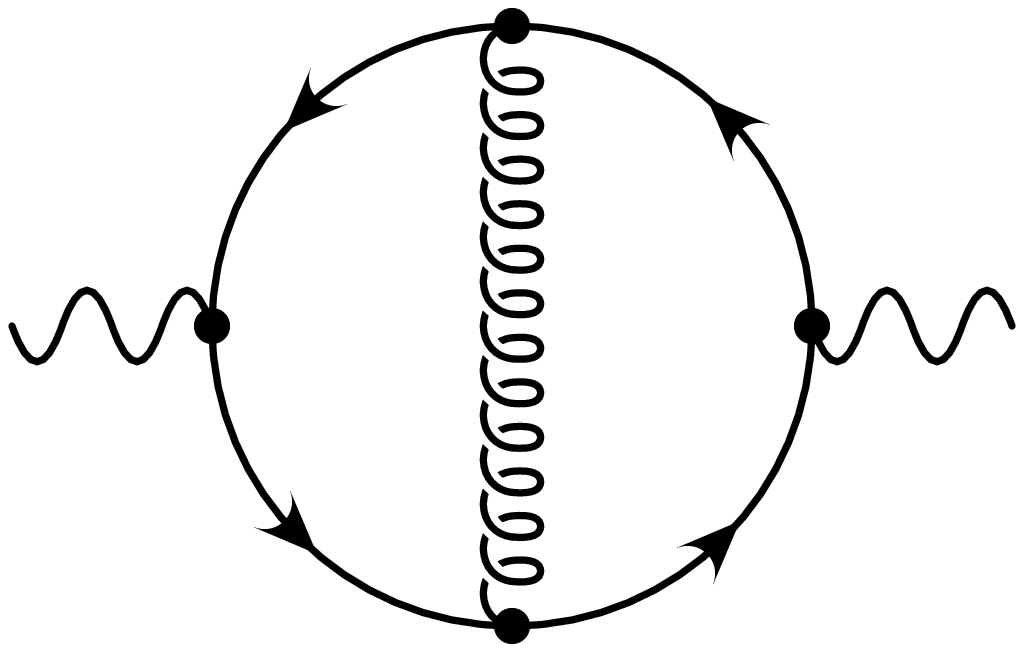} &
\insdia{-1.3em}{8em}{140 265 440 470}{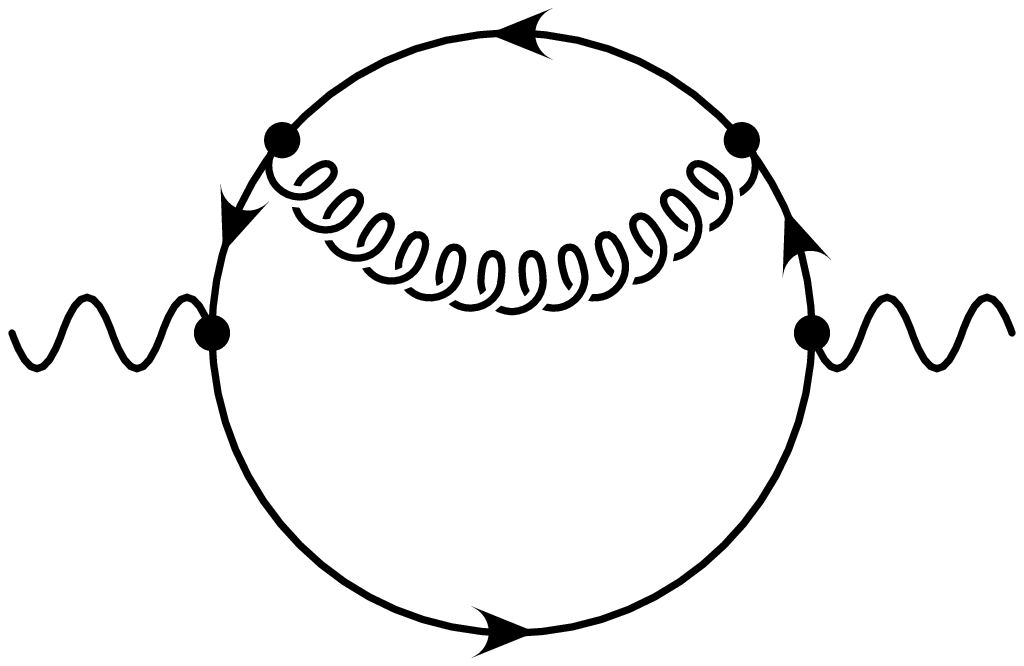}
\end{tabular}
\caption[]{\label{fig::pol} Diagrams contributing to the photon
  polarization function $\Pi(q)$. The outer line denotes a photon of
  momentum $q$, the plain lines are quarks and the spiral ones denote gluons.}
\end{center}
\end{figure}

The two classes of single-scale diagrams mentioned above already have a
huge number of important applications.  The most popular one probably is
the total cross section for hadron production in $e^+e^-$ annihilation
(see, e.g., \cite{CKK:review}), usually written as the hadronic $R$
ratio, in the limit of vanishing quark masses. Using the optical
theorem, it can be expressed through the imaginary part of the photon
polarization function:
\begin{subequations}
\begin{align}
& &&R(s) = 12\pi\Im\Pi(q^2)\bigg|_{q^2=s+i\epsilon}\,,& \label{eq::rimpi}\\
&\mbox{where}&&\Pi(q^2) = {-g_{\mu\nu}+q_\mu q_\nu/q^2\over q^2(D-1)}
\Pi_{\mu\nu}(q)\,,&\label{eq::trans}\\
&\mbox{and }&&
\Pi_{\mu\nu}(q) = i\int\dd^4x e^{iq\cdot x} \langle 0|{\rm T} j_\mu(x)
j_\nu(0)|0\rangle\,,&j_\mu = \bar\psi\gamma_\mu\psi\,.\label{eq::pimunu}
\end{align}
\end{subequations}
$\psi$ is a quark field of mass $m$.  The diagrams contributing to
$\Pi_{\mu\nu}(q)$ up to two-loop order are shown in Fig.~\ref{fig::pol}.
Up to three-loop order, such diagrams can directly be computed by the
program {\tt MINCER} mentioned before ({\it nota bene} in the massless
limit!). Another important application is the computation of moments of
the polarization function, ${\partial^n/\partial (q^2)^n}
\cdot\Pi(q^2)|_{q^2 = 0}$.  They can be obtained by applying the
derivatives and the nullification of $q^2$ {\it before} performing the
loop integrations~\cite{CheKueSte96}. Furthermore, since renormalization
group functions like the QCD $\beta$ function or anomalous dimensions in
the $\msbar$ scheme are independent of any masses and momenta, their
evaluation can be performed by computing single-scale diagrams.

However, returning to the $R$ ratio defined above, the limit of
vanishing quark mass may not be satisfactory, especially if one is
interested in energy regions not too far above one of the quark
thresholds. As long as the exact evaluation of three-loop diagrams
involving a non-vanishing mass as well as an arbitrary external momentum
is not possible, one may hope to reduce the integrals to single-scale
diagrams by performing an expansion in the quark mass. In the optimal
case, a finite number of terms in the expansion will approximate the
full result to reasonable accuracy, and the inclusion of higher order
terms will gradually decrease the error.  To get an idea on what the
result should look like, let us consider the exact one-loop result for
the photon polarization function:
\begin{equation}\label{eq::pi0vexact}
\begin{split}
& \Pi^{(0)}(q^2) = {3\over 16\pi^2}\left( {4\over 3\epsilon} + 
  {20\over 9} + {4\over 3}\,\logmum + {4\over 3 z} -
{4(1-z)(1+2z)\over 3z} G(z)\right)\,,\\
& \mbox{where}\\
&  G(z) = {2u\ln u\over u^2-1}\,, \quad
  u = {\sqrt{1-1/z}-1\over \sqrt{1-1/z}+1}\,,\quad z = {q^2\over 4
    m^2}\,,\quad \logmum = \ln{\mu^2\over m^2}\,,
\end{split}
\end{equation}
and $\mu$ is the renormalization scale.  The pole\footnote{The
  accompanying $\ln 4\pi$ and $\gamma_{\rm E}$ are suppressed throughout
  the paper.}  in $\epsilon = (D-4)/2$, where $D$ is the space-time
dimension, will eventually disappear upon global renormalization (e.g.,
by requiring $\Pi(0) = 0$). The result of (\ref{eq::pi0vexact}) can be
expanded in terms of small mass $m$, yielding:
\begin{equation}\label{eq::pi0v}
  \begin{split}
  {\Pi}^{(0)}(q^2) &= {3 \over 16 \pi^2}\, \bigg\{
   {4\over 3\epsilon} +
  {20\over 9}
  - {4\over 3}\,\logqmu\\&
  + 8\,{{m}^2\over q^2}
  + \left({{m}^2\over q^2}\right)^{2} \, \bigg[
  4
  + 8\,\logqmms
  \bigg]\bigg\} + \ldots\,,
\end{split}
\end{equation}
with $\logqmu = \ln(-q^2/\mu^2)$ and $\logqmms = \ln(-q^2/m^2)$. One
observes that the coefficients of the series in $m^2/q^2$ contain
non-analytical pieces in terms of logarithms. They develop an imaginary
part by means of
\begin{equation}
\ln(-s - i\epsilon) = -i\pi + \ln {s}\,\qquad
\mbox{for } s>0.
\end{equation}
The hadronic $R$ ratio is therefore given by (see (\ref{eq::rimpi})):
\begin{equation}
R(s) = 3\left(1-6\left({m^2\over s}\right)^2 + \ldots\right)\,.
\end{equation}

The question is now if it is possible to obtain the expansion given in
(\ref{eq::pi0v}) directly from the Feynman integrals, i.e.\ without
having to know the exact result.  As a first guess one may try to
perform a Taylor expansion of the integrand, thereby arriving at
massless propagator diagrams.  However, it is clear that such a ``naive
Taylor expansion'' can not be the whole answer. For example, it is
impossible to reproduce the logarithmic mass dependence in this way.
Nevertheless, let us look at the result:
\begin{equation}
\begin{split}
{\cal T}_{m}\Pi^{(0)} &= 
{3\over 16\pi^2}\bigg\{ {4\over 3\epsilon} + {20\over 9} - {4\over 3}\logqmu
\\&
+ 8\,{m^2\over q^2} + \left({m^2\over q^2}\right)^2\bigg[-{8\over \epsilon} -
8 + 8\,\logqmu\bigg]\bigg\} + \cdots\,.
\label{eq::pi0exp}
\end{split}
\end{equation}
In fact, the first two orders in $m^2/q^2$ are reproduced correctly. The
$m^4/q^4$ term, however, is completely different, and there is even an
additional pole in $\epsilon$.  Only the logarithmic $q^2$ dependence is
reproduced. Thus, in general the naive Taylor expansion is not
sufficient to arrive at the desired result. However, in the next section
we will see that by including well-defined additional terms one indeed
can obtain the correct expansion.


\section{Asymptotic Behavior}\label{sec::asym}


This section is divided into three parts, all concerned with the problem
of expanding Feynman diagrams in their external parameters, as it was
raised in Section~\ref{sec::status}.  Cross-references between the three
parts of this section will demonstrate the close correspondence of the
individual formulations.

In Section~\ref{sec::ope}, the problem will be approached from a field
theoretical point of view. The resulting expansion will be derived from
the operator product expansion formulated in the $\msbar$ scheme.  The
viewpoint of Section~\ref{sec::regions}, on the other hand, examines the
individual Feynman integrals that contribute to a certain problem. By a
thorough investigation of the integration regions for the loop momenta
and a subsequent Taylor expansion in the appropriate variables, one can
derive rules that allow to obtain the expansion of the full result in a
very efficient way.

In certain cases these rules could be phrased in a mainly diagrammatical
language. For diagrams involving large external momenta or large masses
this graphical formulation will be described in
Section~\ref{sec::lmphmp}.


%
\subsection{Operator Product Expansion}\label{sec::ope}
The asymptotic behavior of the two-point correlator of
(\ref{eq::pimunu}) in the limit $-q^2=Q^2\to\infty$ is formally known to
all orders of perturbation theory. It is given by an operator product
expansion (OPE):
\begin{equation}
\Pi_{\mu\nu}(q) = i\int\dd^4x e^{iq\cdot x} \langle 0|{\rm T} j_\mu(x)
j_\nu(0)|0\rangle
\stackrel{-q^2\to \infty}{\rightarrow} \sum_n C_{n,\mu\nu}
\langle\opo_n\rangle\,.
\end{equation}
The $C_{n,\mu\nu}$ are complex functions, and the $\opo_n$ are operators
composite of fields of the QCD Lagrangian. 
We keep only Lorentz
scalar operators because all others vanish when sandwiched between the
vacuum states. 
Transversal coefficient functions $C_n$ will be defined in analogy to
the Eq.~(\ref{eq::trans}).
The operators are usually sorted according to their mass
dimension. It is convenient to allow only operators of even mass
dimension which is achieved by appropriate factors of the quark mass $m$
(only one quark shall be considered as massive for the sake of
clarity).
Up to dimension four, the following set of operators is relevant:
\begin{eqnarray}
&&\opo^{(0)}= {\bf 1}\,,\qquad \opo^{(2)} = {\bf m}^2\,,
\label{eq::opo}
\\
&&\opo^{(4)}_1 = G_{\mu\nu}^2\,,\quad \opo^{(4)}_2 = m\bar\psi\psi\,,\quad
\opo^{(4)}_3 = {\bf m}^4\nnb\,,
\end{eqnarray}
where $G_{\mu\nu}$ is the gluonic field strength tensor and $\psi$ is
again the quark field (the superscript ``(4)'' of the dimension-4
operators will be dropped in what follows).  If the operators are
understood to be normal ordered, the vacuum expectation values of the
non-trivial (i.e.~not proportional to unity) operators are equal to zero
in perturbation theory. In such an approach these operators are used to
parameterize non-perturbative effects (see, e.g., \cite{rueckl:ustron}).

\begin{figure}
  \begin{center}
    \begin{tabular}{ccc}
      $\opo_1$ & $\opo_2$ & $\opo_2$\\
  \leavevmode \epsfxsize=\fsize \epsffile[180 265 400
  470]{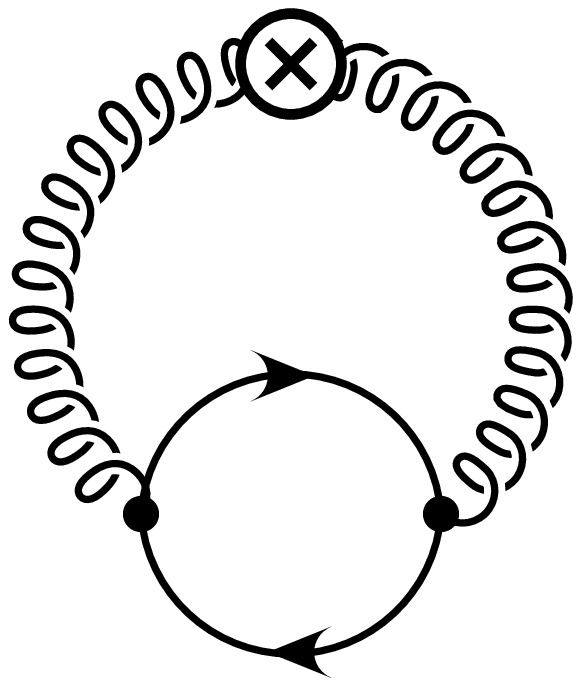} &
  \epsfxsize=\fsize \epsffile[180 265 400 470]{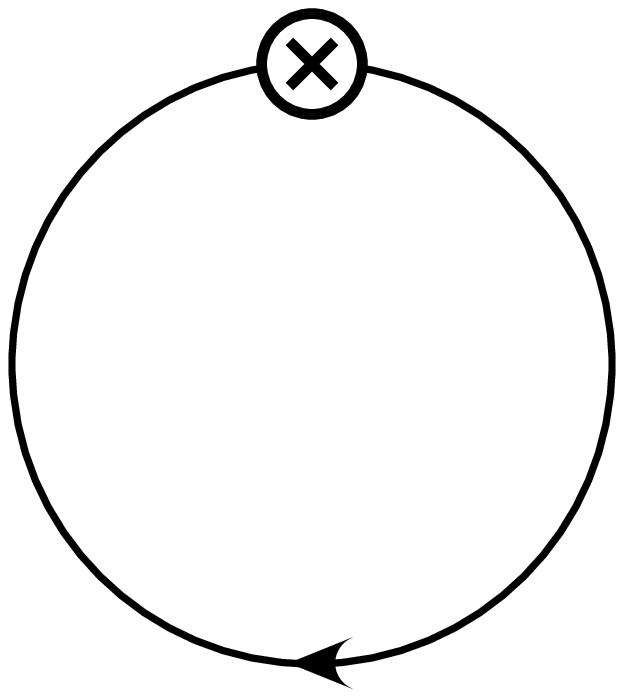} &
  \epsfxsize=\fsize \epsffile[180 265 400 470]{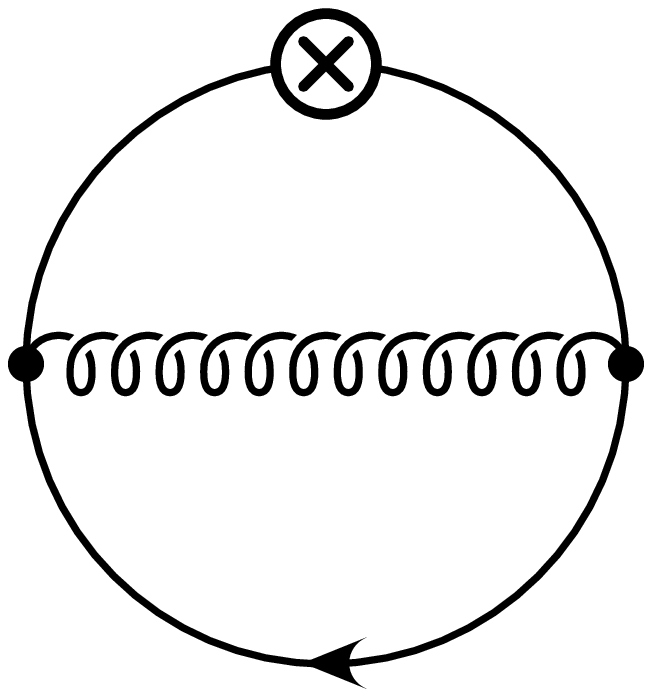}
\end{tabular}
\caption[]{\label{fig::vac} Sample diagrams contributing to $\langle
  \opo_1\rangle$ and $\langle\opo_2\rangle$.}
\end{center}
\end{figure}

On the other hand, if one abandons normal ordering and applies minimal
subtraction, the vacuum expectation values of the field operators
receive also perturbative contributions. The corresponding diagrams are
massive tadpoles which by definition only depend on the quark mass and
the renormalization scale $\mu$. Examples are shown in
Fig.~\ref{fig::vac}.  They lead to the following results:
\begin{equation}
\begin{split}
 \la {\bf 1}\ra &= 1\,,\quad
 \la {\bf m}^2 \ra = m^2\,,\quad
 \la {\bf m}^4\ra = m^4\,,\\
\langle\opo_1^\bare\rangle &= 
{3\over 16\pi^2}\,m^4\,\bigg\{{\alpha_s\over \pi}\,\bigg[
{2\over \epsilon^2} 
+ {1\over \epsilon}\left({14\over 3} + 4\,\logmum\right)\\
&+ 10 + 2\,\zeta_2 + {28\over 3}\,\logmum + 4\,\logmum^2\bigg] +
\cdots\bigg\}\,,
\\
%
\langle\opo_2^\bare\rangle &= {3\over 16\pi^2}\,m^4\,\bigg\{
{4\over \epsilon} 
+ 4 
+ 4\,\logmum + {\alpha_s\over \pi}\big[\cdots\big] + \cdots\bigg\}\,.
%
\end{split}
\label{eq::vacoi}
\end{equation}
Only the first non-vanishing order in $\alpha_s$ is quoted here.
However, the results for the operators that are proportional to unity
(${\bf m}^{2n}, n=0,1,2$) are valid to all orders of perturbation theory
by definition.  Note that there are still poles in $\epsilon$ which is
why the operators are marked with the superscript $\bare$. These poles
disappear upon global renormalization, thereby inducing mixing of the
operators according to
\begin{equation}
\opo_n = \sum_m Z_{nm}\opo_m^\bare\,.
\label{eq::opoB}
\end{equation}
For the dimension-4 operators the renormalization matrix $Z_{nm}$ in the
$\msbar$ scheme was computed in \cite{CheSpi87} by expressing it in
terms of the charge and mass renormalization constants of QCD, plus the
one for the QCD vacuum energy. They are currently known to
$\order{\alpha_s^4}$ \cite{beta4,gammam4} and $\order{\alpha_s^3}$
\cite{Che:gamma0}, respectively. The results for the renormalized vacuum
expectation values of $\opo_1$ and $\opo_2$ up to $\order{\alpha_s}$ can
be found in \cite{CheKue94}.

The crucial observation about using the $\msbar$ scheme was made in
\cite{CheGorTka82}: It appears that in this approach the coefficient
functions are independent of the quark masses. They only depend on the
external momentum $q$ and the renormalization scale $\mu$.  It was shown
that their computation can be reduced to the evaluation of massless
propagator-type diagrams.  The most refined method for this purpose is
called the ``method of projectors'' \cite{methproj}: let us define
``bare'' coefficient functions through
\begin{equation}
\sum_nC_{n,\bare}\opo_n^\bare \equiv \sum_n C_n\opo_n\,.
\end{equation}
Then the ones up to dimension four, for example, are obtained in the
following way (only sample diagrams are displayed here)~\cite{SurTka90}:
\begin{equation}
\begin{split}
  [C_\bare^{(0)},&\,\,C_\bare^{(2)},\,\,C_{3,\bare}] =\\
  &=[1,\,\,\doverd{m_\bare^2},\,\,{1\over 2}{\partial^2\over \partial
    (m_\bare^2)^2}] \left[ \insdia{-1.3em}{5em}{140 265 440
      470}{./d1q.ps} + \insdia{-1.3em}{5em}{140 265 440
      470}{./d2q1.ps} + \cdots
  \right]_{m_\bare=0}\,,\\
  C_{1,\bare} &= {\cal P}_1\left(\doverd{p},\doverd{m_\bare}\right)
  \left[\insdia{-1.3em}{5em}{140 265 440
    470}{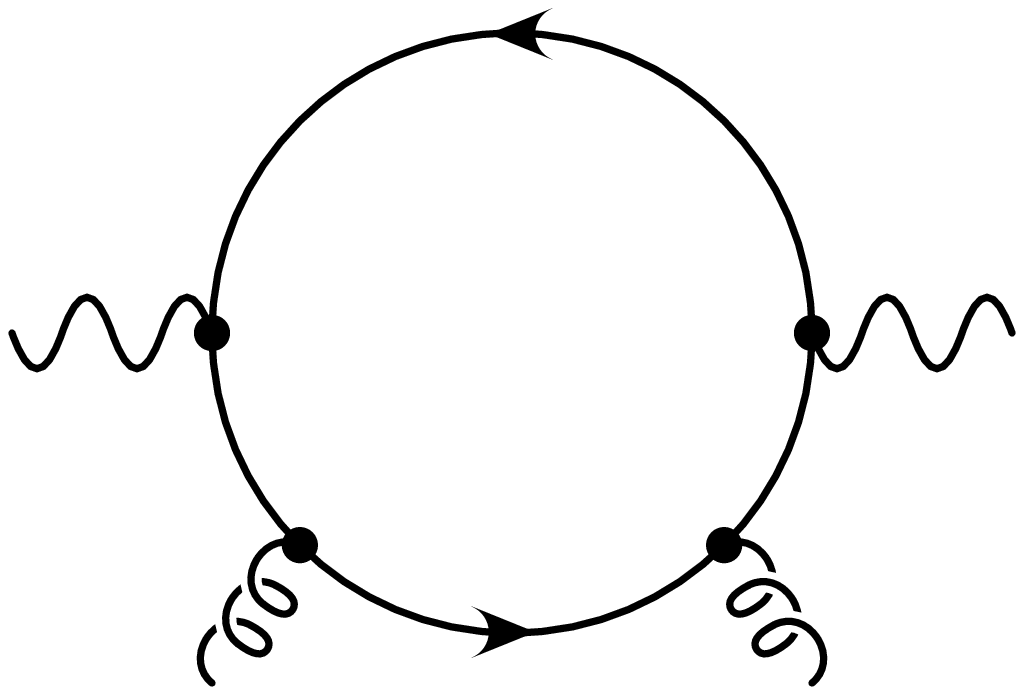} + \cdots\right]_{m_\bare=p=0}\,,\\
  C_{2,\bare} &= {\cal P}_2\left(\doverd{p},\doverd{m_\bare}\right)
  \left[\insdia{-1.3em}{5em}{140
        265 440 470}{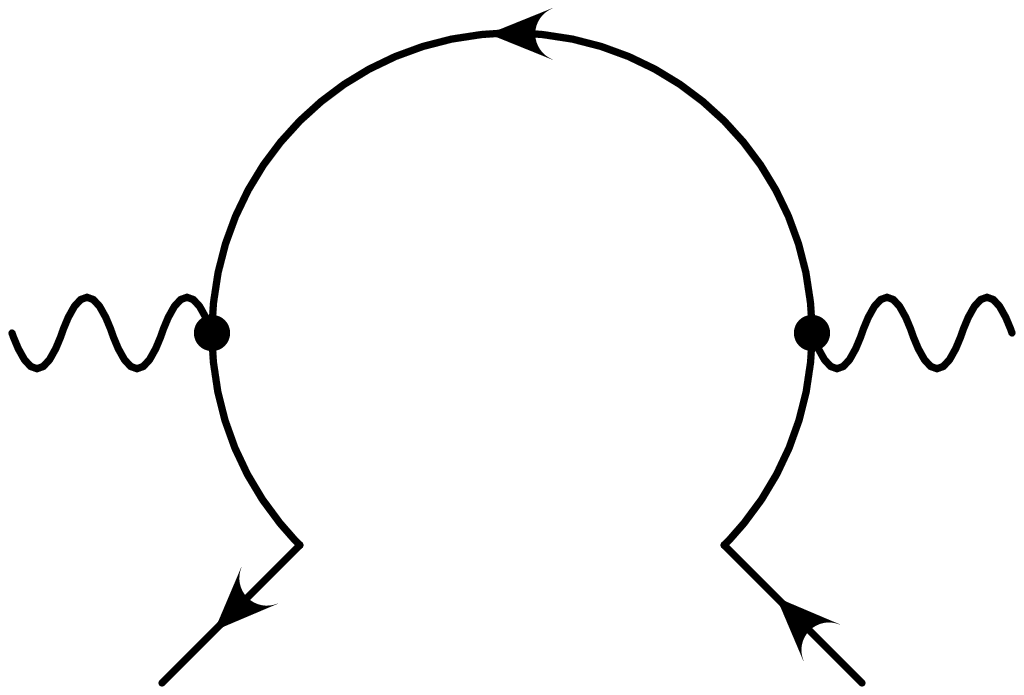} + \cdots + \insdia{-1.3em}{5em}{140 265 440
        470}{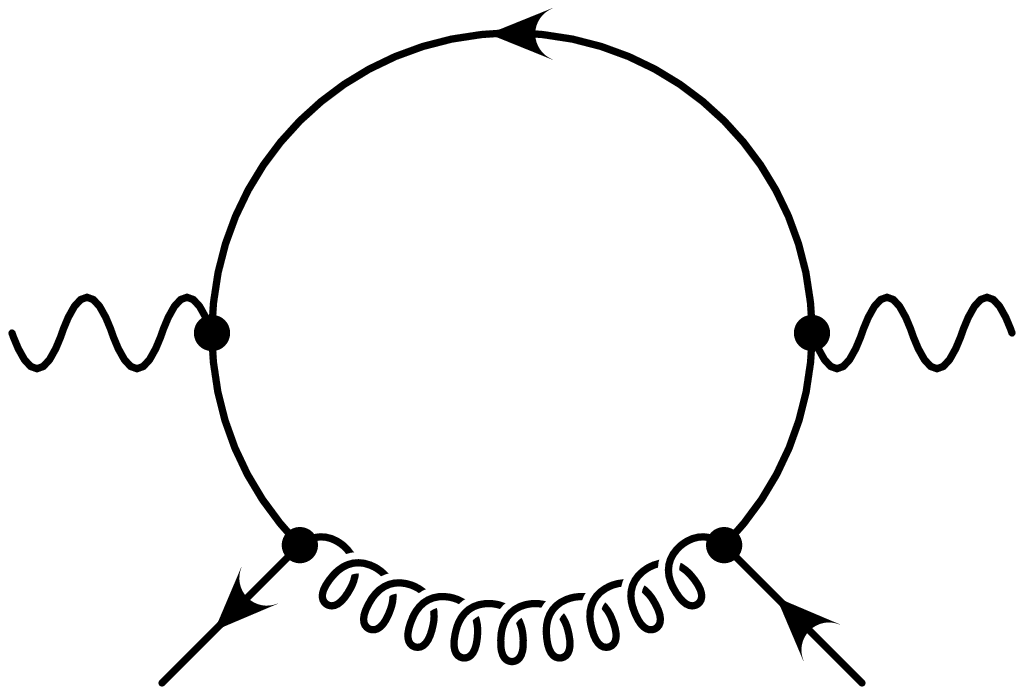} + \cdots\right]_{m_\bare = p = 0}\,,
\end{split}
\label{eq::proj}
\end{equation}
where $p$ is the momentum carried by the external quark-, gluon-, and
ghostlines (the latter arise only at higher orders in $\alpha_s$).  ${\cal
  P}_1$ and ${\cal P}_2$ are ``projectors'' depending polynomially on
the derivatives w.r.t.\ $p$ and $m_\bare$.  For example,
\begin{equation}
{\cal P}_2\left(\doverd{p},\doverd{m_\bare}\right)[\cdots] = 
{1\over 4 n_c}{\rm Tr}\left(\doverd{m_\bare} +
{1\over D}\gamma^\nu\doverd{p^\nu}\right)[\cdots]\,.
\label{eq::p2}
\end{equation}
It is understood that the derivatives act on the {\em integrands} and
the nullification of $p$ and $m_\bare$ is performed {\em before}
integration. Note, however, that the momentum carried by the
external currents (wavy lines), is $q\neq 0$.  Obviously,
$C_\bare^{(0)}$, $C_\bare^{(2)}$, and $C_{3,\bare}$ are just the
coefficients of $m_\bare^0$, $m_\bare^2$, and $m_\bare^4$ of the naive
Taylor expansion, respectively.  The expressions for $C_{1,\bare}$ and
$C_{2,\bare}$, on the other hand, read as follows:
\begin{equation}
\begin{split}
  C_{1,\bare} &= {1\over q^4}\bigg\{{\alpha_s\over
      \pi}\bigg[{1\over 12} + \epsilon\,\left(  {7\over 72} - {1\over
      12}\,\logqmu\right)\bigg] + \ldots\bigg\}\,,\\
  C_{2,\bare} &= {1\over q^4}\bigg\{2 + \epsilon 
    +{\alpha_s\over \pi}\bigg[{2\over 3} +\epsilon\,\left({5\over 3} -
      {2\over 3}\,\logqmu\right)\bigg]
    + \ldots\bigg\}\,,
\end{split}
\label{eq::c1c2b}
\end{equation}
where we have kept the terms up to $\order{\epsilon}$ because they
contribute to the finite part of $\Pi(q^2)$.  In fact, with these
ingredients it is possible to compute the polarization function up to
order $\alpha_s^0 m^4$. Diagrammatically one finds
\begin{equation}
  \insdia{-1.3em}{5em}{140 265 440 470}{./d1q.ps} 
  \stackrel{-q^2\to \infty}{\rightarrow}{\cal T}_{m_\bare}
  \insdia{-1.3em}{5em}{140 265 440 470}{./d1q.ps} 
  +\,\,
  2\,\,\insdia{-1.3em}{5em}{140 265 440 470}{./p2l0.ps} \star
  \insdia{-1.3em}{3.5em}{180 265 400 470}{./vaco2l1.ps}\,,
\label{eq::asym1l}
\end{equation}
where in the second term on the right hand side the projection with
${\cal P}_2$ from (\ref{eq::p2}) is implicit.  The factor 2 arises 
from the symmetrical diagram that also contributes to $C_{2,\bare}$.
Note that there is no
contribution from $\opo_1$ at this order.  The Taylor expansion in the
first term is to be carried out up to $m_\bare^4$. This first term is
given by Eq.~(\ref{eq::pi0exp}), since $m=m_\bare + \order{\alpha_s}$.
Using the results of (\ref{eq::vacoi}) and (\ref{eq::c1c2b}) for
$C_{2,\bare}$ and $\la\opo_2^{\bare}\ra$, one obtains the result for the
second term of (\ref{eq::asym1l}):
\begin{equation}
2\,C_{2,\bare}\la\opo^\bare_2\ra = {3\over 16\pi^2}\,\left({m^2\over
    q^2}\right)^2\,\bigg[
{8\over \epsilon} + 12 + 8\,\logmum\bigg] + \order{\alpha_s}\,.
\label{eq::c2o2}
\end{equation}
Adding it to (\ref{eq::pi0exp}) exactly reproduces the result for
$\Pi(q^2)$ up to $\order{m^4}$ given in (\ref{eq::pi0v}).

The conclusion from these considerations is that in addition to the
naive Taylor expansion of (\ref{eq::pi0exp}) one should include an extra
term, given by $2\,C_{2,\bare}\la\opo_2^\bare\ra$, in order to arrive at
the correct result for the polarization function.  However, the
important point about Eq.~(\ref{eq::asym1l}) is that the original
diagram is reduced to single-scale factors (i.e., massless propagators
and massive tadpoles).


\subsection{Strategy of Regions\cite{regions,Smi99}}\label{sec::regions}
Let us for the moment forget about OPE again and consider only the
Feynman integral for the one-loop diagram, concentrating on the scalar
case for the sake of clarity:
\begin{equation}
\insdia{-1.3em}{5em}{140 265 440 470}{./d1q.ps}
 \stackrel{\rm scalar}{=} \int \fprop{1}{m}{k}\fprop{1}{m}{(k-Q)}\,,
\end{equation}
where the momenta are taken in Euclidean space and integration is over
$k$. Assume now that $m^2\ll Q^2\, (=-q^2)$. The integral may be split into the
following regions:
\begin{equation}
\begin{array}{llll}
(i):\quad & k^2\gg m^2 \quad&\mbox{and} & (k-Q)^2\gg m^2\\
(ii):\quad & k^2\sim m^2 &\Rightarrow & (k-Q)^2\gg m^2\\
(iii):\quad & (k-Q)^2\sim m^2 &\Rightarrow & k^2\gg m^2\,,
\end{array}
\end{equation}
where $\sim$ means ``of the order of''. In region $(i)$, the integrand
can be expanded in terms of small $m$:
\begin{equation}
\begin{split}
&\int_{(i)}\,\fprop{1}{m}{k}\fprop{1}{m}{(k-Q)} \approx
\int_{(i)}\, {\cal T}_m\fprop{1}{m}{k}\fprop{1}{m}{(k-Q)} =\\
&=\int_{(i)}\, {1\over k^2}{1\over (k-Q)^2}\left(1-{m^2\over k^2} +
  \ldots\right)\left(1-{m^2\over (k-Q)^2} + \ldots\right)\,.
\end{split}
\end{equation}
In region $(ii)$, $k$ is
considered to be of the same order of magnitude as $m$, so one should
expand in $m$ and $k$ at the same time:
\begin{equation}
\begin{split}
  &\int_{(ii)}\,\fprop{1}{m}{k}\fprop{1}{m}{(k-Q)} \approx
  \int_{(ii)}\, {\cal T}_{m,k}\fprop{1}{m}{k}
  \fprop{1}{m}{(k-Q)} =\\
  &=\int_{(ii)}\, {1\over m^2+k^2}{1\over Q^2}
    \left(1-{m^2+k^2-2k\cdot Q\over Q^2} + \ldots\right)\,.
\label{eq::ii}
\end{split}
\end{equation}
Region $(iii)$ can be mapped onto region $(ii)$ by substituting
$k' = Q-k$, meaning $k^{\prime 2}\sim m^2 \Rightarrow (k'-Q)^2\gg m^2$,
and we arrive at an expression analoguous to (\ref{eq::ii}).

Now we add and subtract the complementary regions to the integrals
above and find:
\begin{equation}
\begin{split}
  & \int\,\fprop{1}{m}{k}\fprop{1}{m}{(k-Q)} \approx\\
  & \approx \int\, {\cal T}_{m}\fprop{1}{m}{k}\fprop{1}{m}{(k-Q)} +
  2\int\, \fprop{1}{m}{k}{\cal T}_{m,k}\fprop{1}{m}{(k-Q)}\\
  &\qquad - C\,,
  \label{eq::expc}
\end{split}
\end{equation}
where
\begin{equation}
\begin{split}
C &= \int_{(ii)\cup (iii)} {\cal T}_m \fprop{1}{m}{k}\fprop{1}{m}{(k-Q)}
+ \int_{{(i)\cup (iii)}} \fprop{1}{m}{k}\times
\\&\quad\times{\cal T}_{m,k}\fprop{1}{m}{(k-Q)}
+ \int_{(i)\cup (ii)}\fprop{1}{m}{{k^\prime}} {\cal T}_{m,k'}
\fprop{1}{m}{(k'-Q)}\,.
\end{split}
\end{equation}
In each of the different regions, one can again expand w.r.t.\ the
appropriate parameters, for example:
\begin{equation}
\begin{split}
&\int_{(ii)} {\cal T}_m \fprop{1}{m}{k}\fprop{1}{m}{(k-Q)} \approx\\
&\qquad
\int_{(ii)} \left({\cal T}_m \fprop{1}{m}{k}\right)
\left({\cal T}_{m,k} \fprop{1}{m}{(k-Q)}\right)\,.
\end{split}
\end{equation}
Finally, for $C$ one finds:
\begin{equation}
C = 2\int_{(i)\cup (ii)\cup (iii)}  \left({\cal T}_m \fprop{1}{m}{k}\right)
\left({\cal T}_{m,k} \fprop{1}{m}{(k-Q)}\right)\,,
\end{equation}
which corresponds to massless tadpole integrals and therefore vanishes
in dimensional regularization which we are using throughout.\footnote{ I
  acknowledge a useful conversation with K.~Melnikov on this issue.}

The second term on the r.h.s.\ of Eq.~(\ref{eq::expc}) can be
represented graphically as
\begin{equation}
2\,\cdot\insdia{-1.3em}{5em}{140 265 440 470}{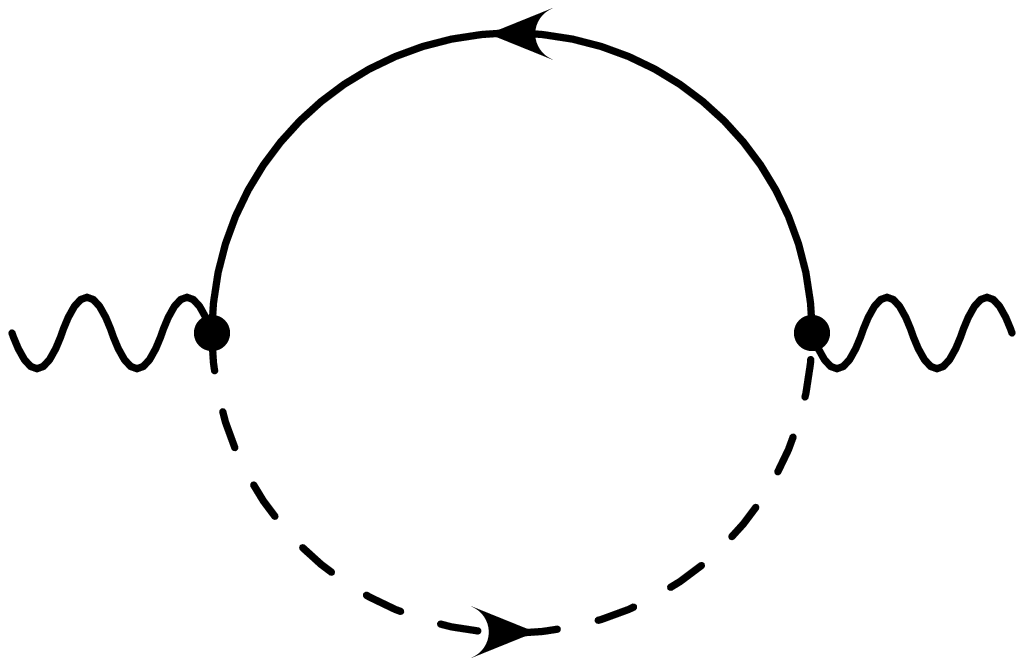} \,,
\label{eq::d1q_2}
\end{equation}
meaning that the solid line should be expanded in all quantities but
$Q$. Doing this, the actual integral will be of the tadpole type, with a
more or less complicated vertex insertion. It is obtained from the
original diagram by shrinking the solid lines shown in (\ref{eq::d1q_2})
to a point.  In this sense, one can again use the diagrammatical
equation (\ref{eq::asym1l}) in order to represent Eq.~(\ref{eq::expc}).
However, the interpretation is slightly different: In the second term,
the diagram left of ``$\star$'' means the solid line of
(\ref{eq::d1q_2}), and the one right of ``$\star$'' is the tadpole which
remains when shrinking the expanded lines to a point.  But the main
difference between both interpretations of Eq.~(\ref{eq::asym1l}) is
that according to Section~\ref{sec::ope} it is only valid up to terms of
order $(m^2/q^2)^2$; according to Section~\ref{sec::regions}, however,
one may obtain the diagram to the r.h.s.\ up to arbitrary high powers in
$m^2/q^2$ by pushing the Taylor expansions on the l.h.s.\ to
sufficiently large order.


\subsection{Large momentum and hard mass
  procedure
  \cite{CheGorTka82,PivTka84,methproj,CheSmi87,Gor89,Smi91}}\label{sec::lmphmp}
So far, two different viewpoints for the small-mass expansion of the
polarization function have been presented. One is based on OPE and uses
the language of quantum fields, while the second one examines the very
Feynman {\em integrals}. The third viewpoint we will focus on is
formulated mainly in terms of Feynman {\em diagrams}. Given a certain
Feynman diagram with a particular distribution of masses and external
momenta, the method generates a set of simpler diagrams which correspond
to the asymptotic form of the original diagram in certain limits of the
external parameters.

The particular case of a large external momentum as it was considered so
far is usually referred to as ``large momentum procedure''. The opposite
case of a mass being much larger than any other scale is called ``hard
mass procedure'' and will be addressed below.

The prescription for the asymptotic expansion of Feynman diagrams can be
summarized by the following formula~\cite{Smi91}:
\begin{eqnarray}
{\cal F}(\Gamma) & \to &
\sum_\gamma {\cal F}(\Gamma\backslash\gamma)
\,\,\star\,\, {\cal T}{\cal F}(\gamma)
\,.
\label{eq::asexp}
\end{eqnarray}
Here, $\Gamma$ is the Feynman diagram under consideration, and ${\cal
  F}(\Gamma)$ is the corresponding Feynman integral.  It shall contain
either a set of large external momenta $\{Q\}$ or of large masses
$\{M\}$.  The arrow ($\to$) denotes that the r.h.s.\ is valid in the
asymptotic limit of the $M$ or $Q$ going to infinity. The sum goes over
all subgraphs $\gamma$ of $\Gamma$ that fulfill certain conditions to be
described below.  $\Gamma\backslash\gamma$ means the diagram that
results when, within $\Gamma$, all lines of $\gamma$ are shrunk to
points.  ${\cal T}$ means Taylor expansion w.r.t.\ all masses and
external momenta that are {\em not} large. In particular, also those
external momenta of $\gamma$ that appear to be integration momenta in
$\Gamma$ have to be considered as small. The Taylor expansions are
understood to be applied before any loop integrations are performed.  In
the following we will refer to the $\gamma$ as {\em hard subgraphs} or
simply {\em subgraphs}, to $\Gamma\backslash\gamma$ as the corresponding
{\em co-subgraphs}. The ``$\star$'' means that ${\cal TF}(\gamma)$ shall
be inserted into ${\cal F}(\Gamma\backslash\gamma)$ at the point to
which $\gamma$ was contracted.

In other words: within $\Gamma$, all propagators of $\gamma$ have to be
expanded w.r.t.\ the masses and external momenta of $\gamma$ that are
not large.


\subsubsection{Large momentum procedure}\label{sec::lmp}
For the particular case of the large momentum procedure, the
conditions that specify the hard subgraphs are as follows: 

Every $\gamma$ has to (i) contain all vertices where a large momentum
enters or leaves the graph and (ii) be one-particle irreducible if these
vertices were connected by an extra line.

As an example, consider again the one-loop diagram contributing to the
photon polarization function, Fig.~\ref{fig::pol}(a), in the limit of large
external momentum.
The set of hard subgraphs that emerge consists of
three diagrams: first there is the diagram itself; the corresponding
co-subgraph is just a point.  The second subgraph is the one shown in
(\ref{eq::d1q_2}) if the dashed line is omitted, and the third subgraph
is the one symmetrical to that. The corresponding co-subgraphs are one-loop
tadpole diagrams.
In this way one again arrives at Eq.~(\ref{eq::asym1l}). The
interpretation of the terms is the same as it was in the approach
of Section~\ref{sec::regions}.

As a two-loop example, let us examine the diagram shown in
Fig.~\ref{fig::pol}~(b):
\begin{equation}
\begin{split}
  \insdia{-1.3em}{5em}{140 265 440 470}{./d2q1.ps}
 & \stackrel{-q^2\to \infty}{\rightarrow}
  \insdia{-1.3em}{5em}{140 265 440 470}{./d2q1.ps} \star 1 
\\&+\,\,
  4\,\,\insdia{-1.3em}{5em}{140 265 440 470}{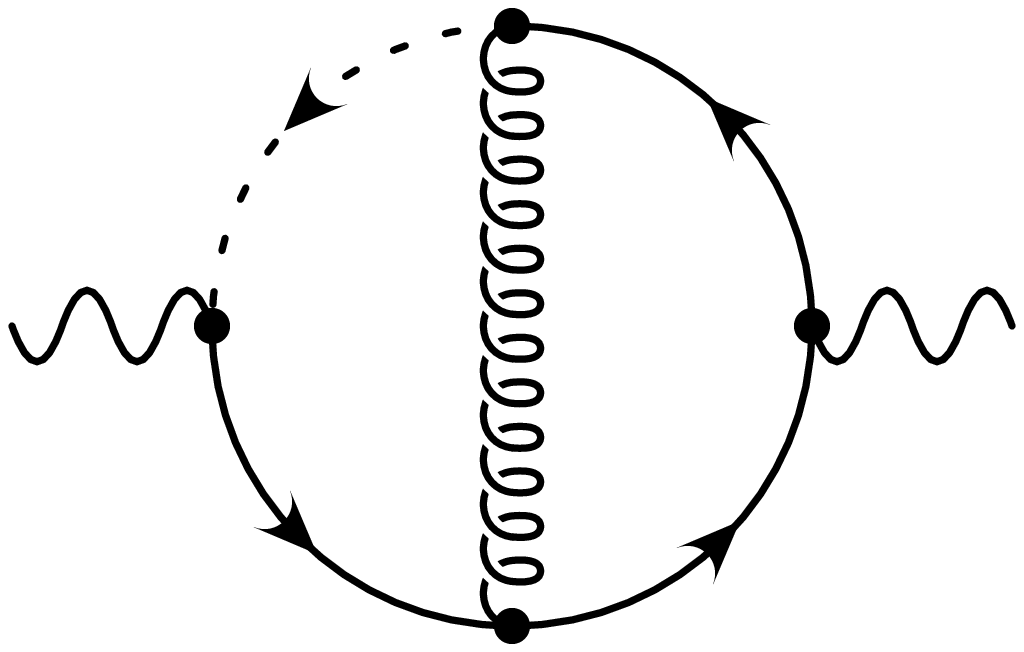} \star
  \insdia{-1.3em}{3.5em}{180 265 400 470}{./vaco2l1.ps}
  +\,\,
  2\,\,\insdia{-1.3em}{5em}{140 265 440 470}{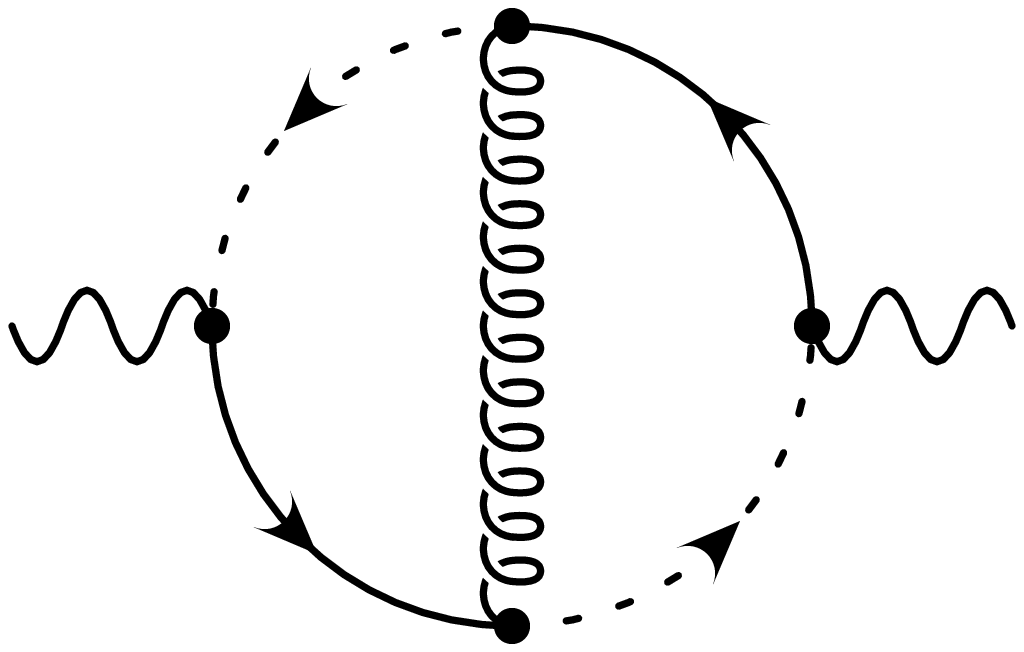} \star
  \insdia{-1.3em}{3.5em}{180 265 400 470}{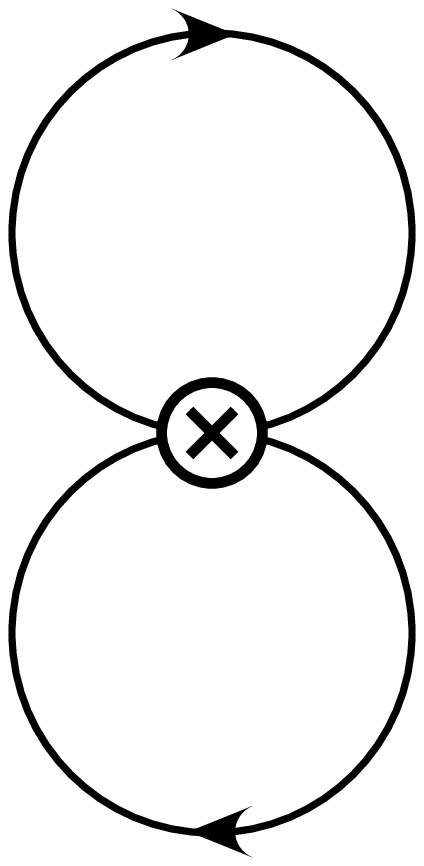}\,
  \\&+\,\,
  2\,\,\insdia{-1.3em}{5em}{140 265 440 470}{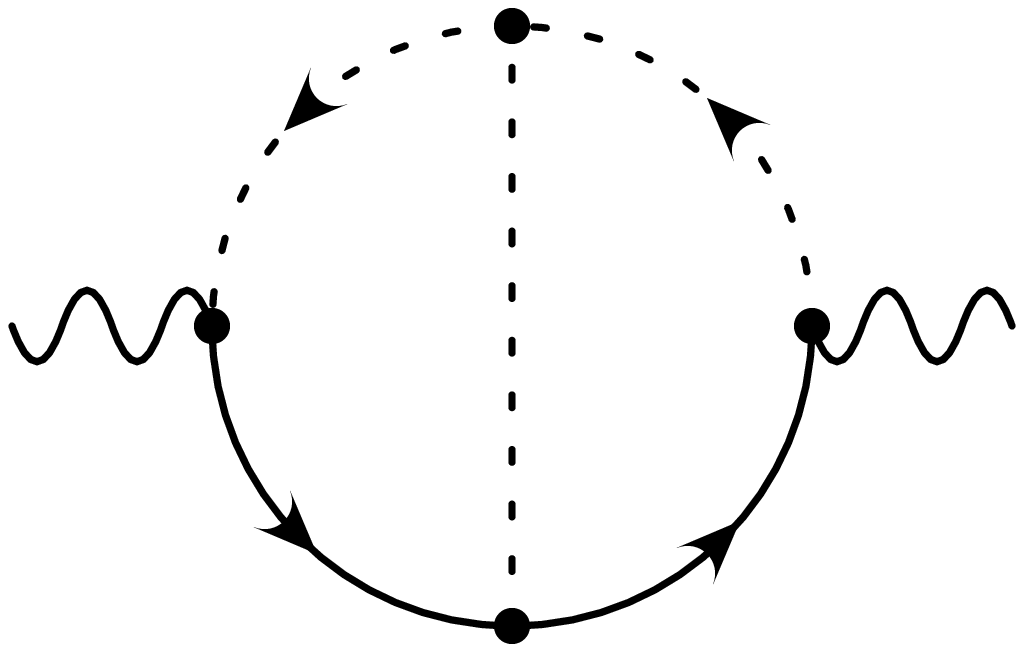} \star
  \insdia{-1.3em}{3.5em}{180 265 400 470}{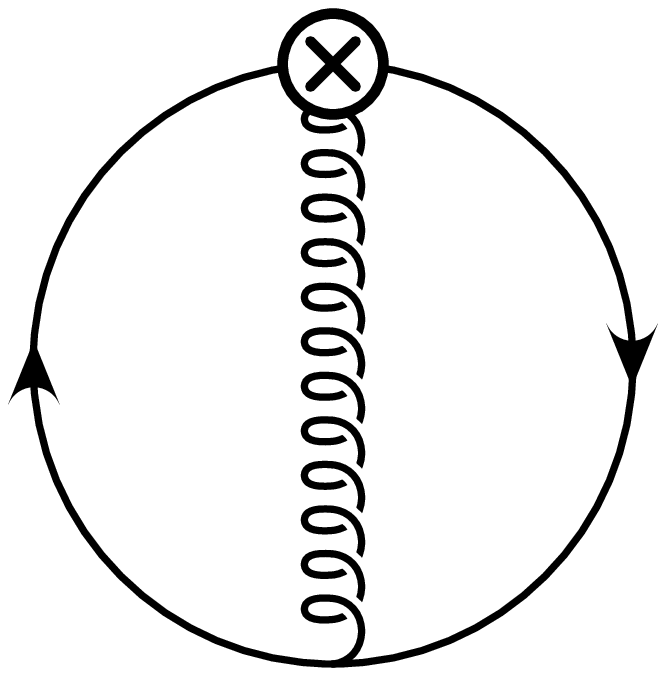}\,
  +\,\,
  \insdia{-1.3em}{5em}{140 265 440 470}{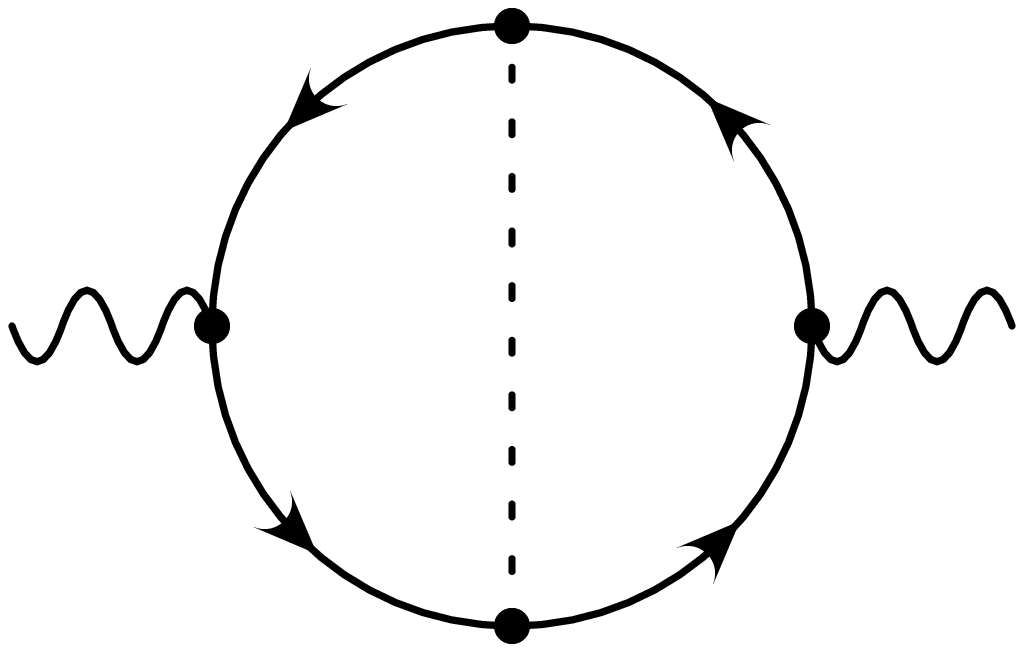} \star
  \insdia{-1.4em}{3.5em}{180 265 400 470}{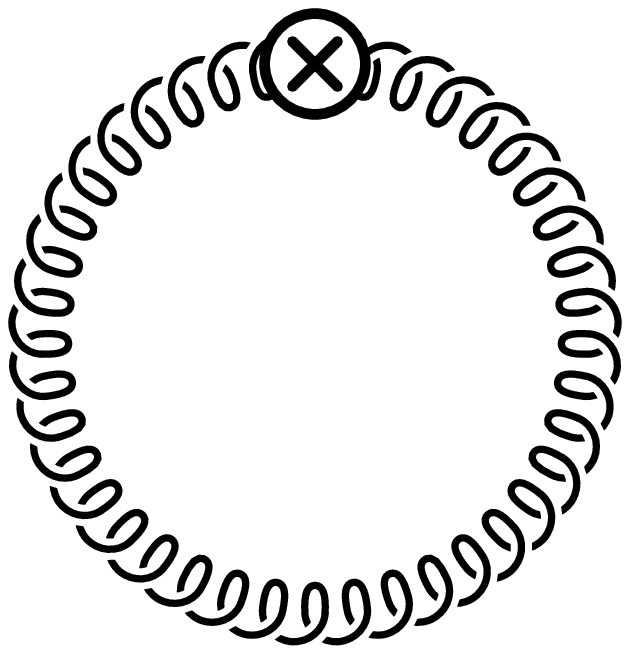}\,,
\end{split}
\label{eq::asym2l}
\end{equation}
where the subgraphs are given by the solid lines of the diagrams left of
``$\star$''.  The first term on the r.h.s.\ corresponds to the naive
Taylor expansion. Note that the last diagram is zero in dimensional
regularization.

It is interesting to relate this set of terms to the viewpoints of the
previous section. In the operator language, the first term corresponds
to the trivial operators ${\bf m}^{2n}$. The second one (actually only
its $m^4/q^4$ contribution) is related to
$C_{2,\bare}\la\opo_2^\bare\ra$, the last one to
$C_{1,\bare}\la\opo_1^\bare\ra$.  The other two terms, however, have no
correspondence to any of the operators of Section~\ref{sec::ope}. This
is due to the fact that we considered only operators up to dimension
four. But relation (\ref{eq::asym2l}) is valid up to arbitrary orders in
$m^2/q^2$. So the conclusion is that the third and fourth term on the
l.h.s.\ are of order $(m^2/q^2)^3$ or higher.

The viewpoint described in Section~\ref{sec::regions}, on the other
hand, reproduces exactly the same terms as shown above.


\subsubsection{Hard mass procedure}\label{sec::hmp}
So far only the case of an external momentum being much larger than any
other scale of the problem was considered.  In this section the
so-called hard mass procedure will be discussed. Here it is assumed that
the diagram carries a mass that is much larger than all other masses and
external momenta. Equation (\ref{eq::asexp}) remains valid, only the
classification of the subgraphs is different. In the case of the hard
mass procedure, $\gamma$ must (i) contain all lines carrying a large
mass (ii) be one-particle irreducible in its connected parts after
contracting the heavy lines.  Consider the following two-loop diagram as
an example:
\begin{equation}
\insdia{-3em}{10em}{140 265 440 470}{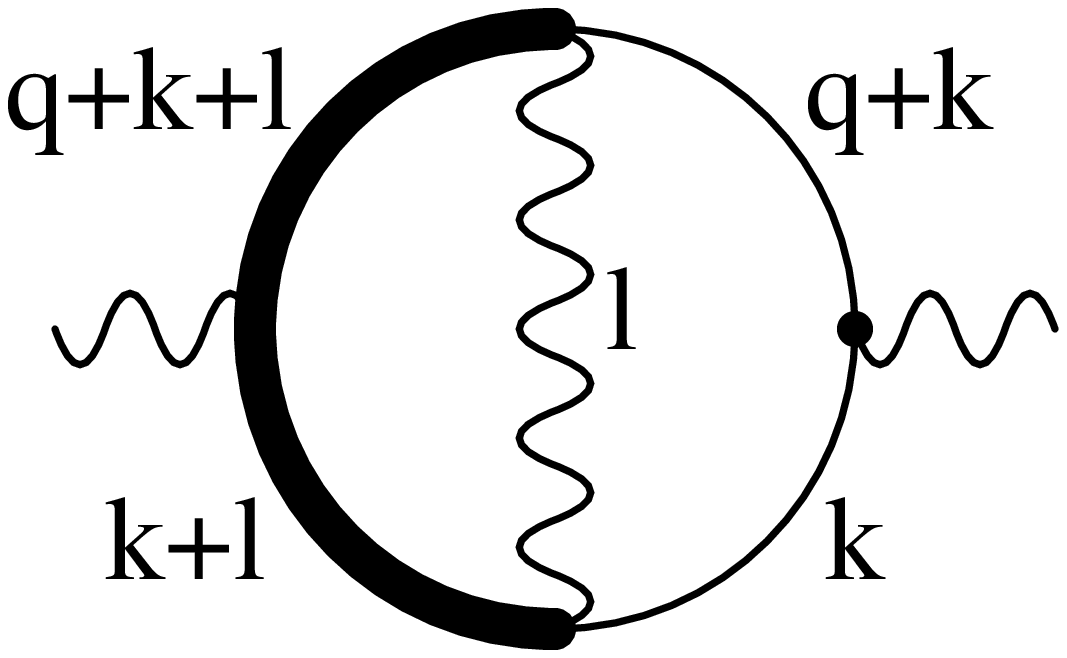} \quad,
\label{eq::d2q1m}
\end{equation}
where $q$ is the external momentum flowing through the diagram from
right to left. The mass of the thick line will be denoted by $M$. The
imaginary part of (\ref{eq::d2q1m}) contributes to the electro-weak
one-loop corrections of the process $Z\to b\bar b$ if top-quarks are
attributed to the thick lines, $b$ quarks to the plain thin lines, $W$
bosons to the inner, and $Z$ bosons to the outer wavy lines (see also
Section~\ref{sec::app} below).
In the limit $q^2\ll M^2$, the following subdiagrams emerge:
\begin{equation}
\begin{tabular}{lllll}
\begin{tabular}{l}
$k^2\sim M^2$\\ $l^2\, \sim  M^2$ 
\end{tabular}
&:& \insdia{-1.4em}{5em}{140 265 440
  470}{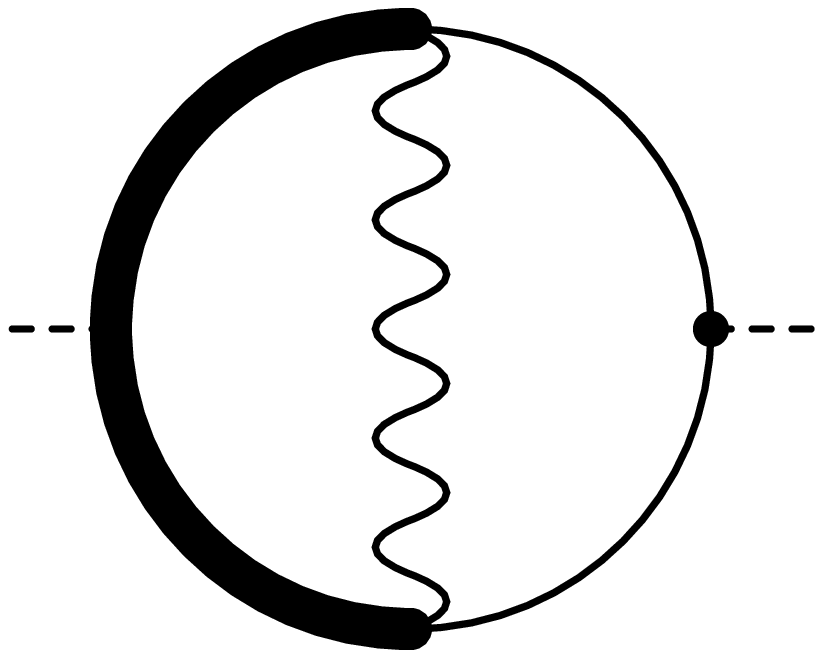}
&$\star$& \insdia{-1.5em}{4em}{180 265 400
  470}{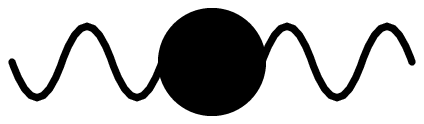}\,, \\
\begin{tabular}{l}
$k^2\sim M^2$\\ $l^2  \ll  M^2$ 
\end{tabular}
&:& \insdia{-1.3em}{5em}{140 265 440
  470}{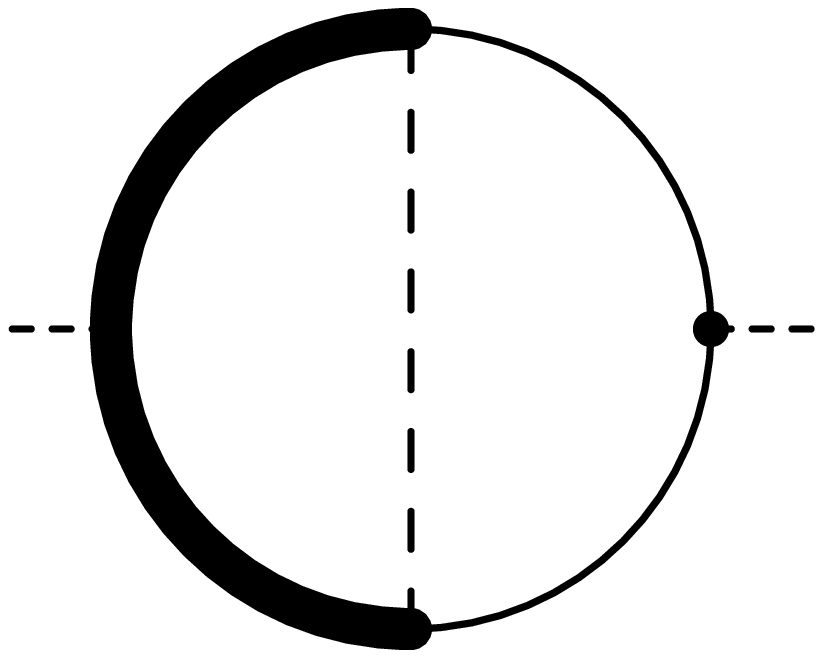}
&$\star$& \insdia{-1.5em}{4em}{180 265 400
  470}{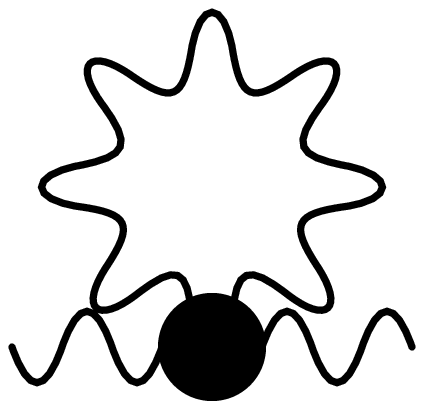}\,, \\
\begin{tabular}{l}
$k^2\ll M^2$\\ $l^2\, \sim\, M^2$ 
\end{tabular}
&:& \insdia{-1.3em}{5em}{140 265 440
  470}{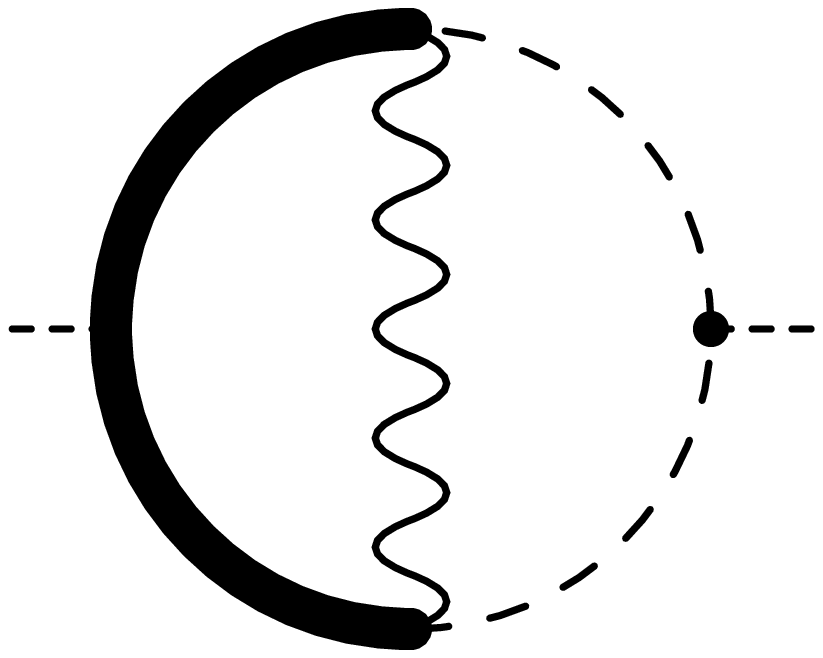}
&$\star$& \insdia{-1.em}{4em}{140 265 440
  470}{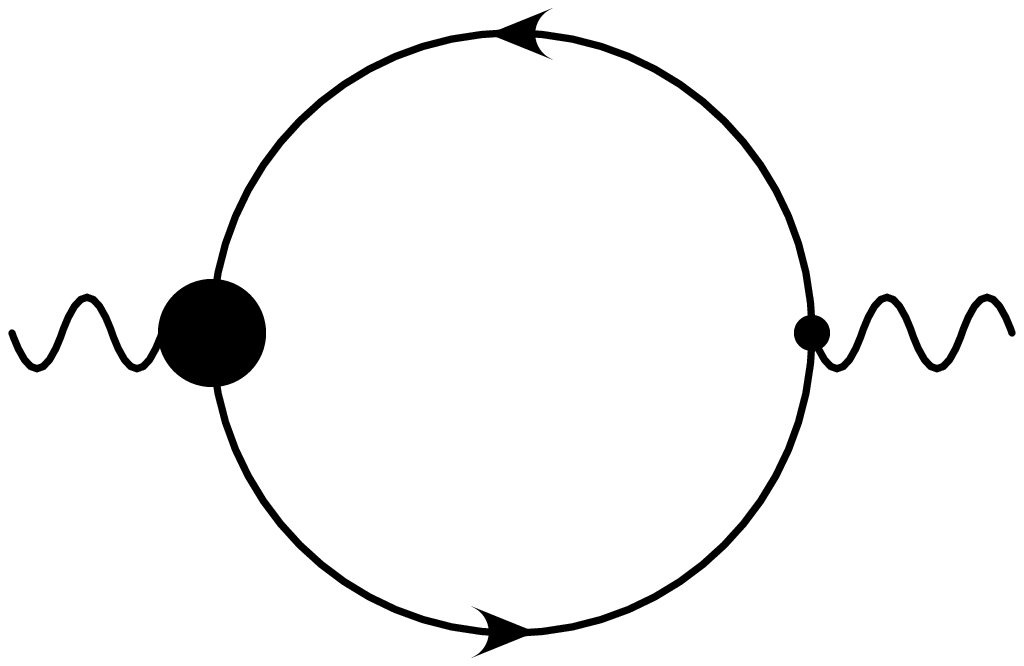}\,, \\%
\begin{tabular}{l}
$k^2\ll M^2$\\ $l^2\,  \ll  M^2$ 
\end{tabular}
&:& \insdia{-1.3em}{5em}{140 265 440
  470}{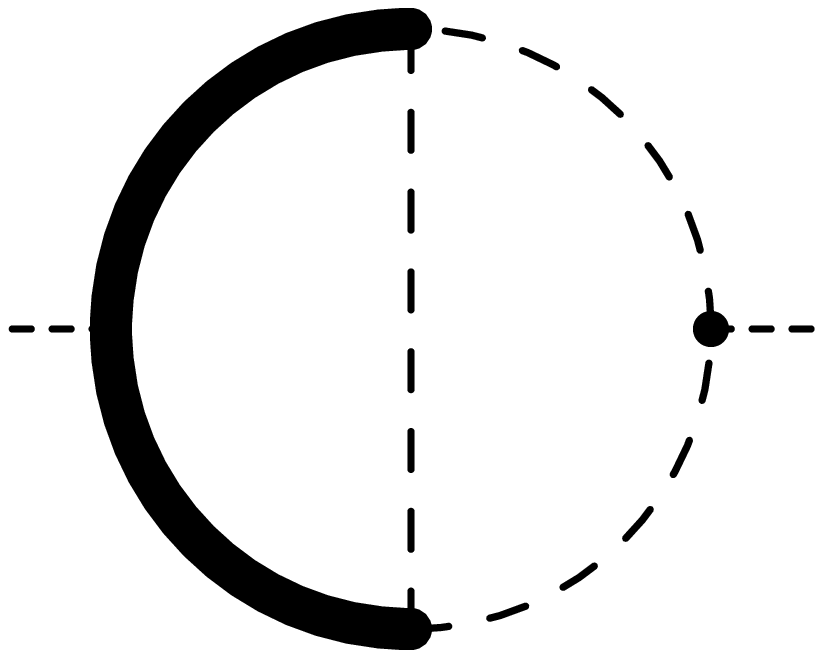}
&$\star$& \insdia{-1.em}{4em}{140 265 440
  470}{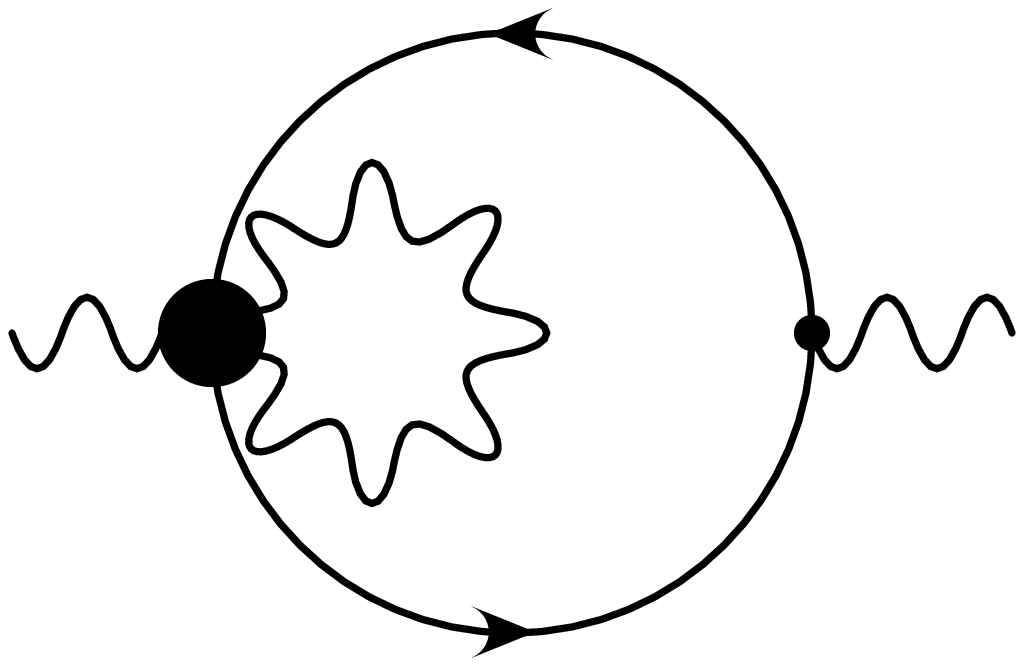}\,, \\
\end{tabular}
\label{eq::hmp}
\end{equation}
where we have indicated the region of loop momenta that generates the
corresponding subdiagram according to the considerations or
Section~\ref{sec::regions}. It is again understood that the solid lines
are to be expanded w.r.t.\ all external momenta and masses except $M$.
The co-subgraphs, shown right of ``$\star$'', are obtained by contracting
the solid lines to points.


\subsubsection{Successive application of large momentum and hard mass
  procedure}
In theories with many different particles of various masses, a realistic
process generally involves several scales. The successive application of
the large momentum and the hard mass procedure can be used to reduce any
Feynman diagram to single-scale factors in this case, as long as a
hierarchy among the different scales can be defined.  This strategy was
followed, for example, in the calculation of the anomalous magnetic
moment of the muon \cite{mug-2}, and later, using an automated setup, to
the decay of the $Z$ boson into $b$ quarks \cite{HSS:zbb} (see
Section~\ref{sec::app} below).


\subsection{Expansions in other limits and range of
  applicability}\label{sec::other}
Two concluding remarks shall be made to complete this formal section.
First it should be noted that asymptotic expansions have been developed
for various limiting cases (see, e.g., \cite{CzaSmi97,regions}). Only the two most straightforward ones have
been described above. Other situations arise, for example, if the
external momentum is either close to a mass or a threshold of the
diagram. It appears that the most convenient way to formulate these
expansions is in terms of the language of Section~\ref{sec::regions}.
Applications will be discussed briefly in the next Section.

The second remark that should be made concerns the applicability of
asymptotic expansions, or in other words, the convergence properties of
the series. By investigating the global structure of the problem, one
often can read off regions of convergence for the function to be
approximated. These regions may extend to values of the expansion
parameter that are way beyond the initially required conditions.

Restrictions on the region of convergence, on the other hand, are mostly
induced by the presence of thresholds.  Thus, it often appears that with
increasing number of loops the range of validity for the result
decreases due to additional thresholds that are absent at lower order of
perturbation theory.



\section{Asymptotic expansions in practical applications}\label{sec::app}
{\bf \bld{R(s)} to \bld{\order{\alpha_s^2}}:} in \cite{CHKS:m12} the
method of Section~\ref{sec::lmp} was directly applied to the three-loop
polarization function.  For this purpose, the diagrammatical
prescriptions were implemented in a computer program, called {\tt
  lmp}~\cite{Har:diss}, in order to cope with the large number of subdiagrams
that had to be generated. For details on the calculation and the
discussion of the results and their convergence properties we refer to
\cite{CHKS:m12,CHKS:lausanne}.

{\bf RG functions in MOM scheme:} the large momentum procedure was also
used to evaluate the relations between the renormalization constants of
QCD in the $\msbar$ scheme to the ones in other schemes, in particular
the MOM scheme, at $\order{\alpha_s^3}$. See \cite{CheRet99} for more
details.

{\bf \bld{Z\to b\bar b} to order \bld{G_{\rm F}\alpha_s}:} a sample
diagram whose imaginary part contributes at $\order{G_{\rm F}}$ is shown
in Eq.~(\ref{eq::d2q1m}). At order $G_{\rm F}\alpha_s$, an additional
gluon has to be attached. One can apply the \hmp\ w.r.t.\ $M_t$ in this
case, and it turns out that the leading term is proportional to $M_t^2$
due to the large mass splitting between the bottom and the top quark.
There are two other scales in the problem, $M_W$ and $M_Z$.
In~\cite{HSS:zbb} they were factorized by a successive application of
the \hmp\ according to $M_W\gg M_Z$, an inequality whose use clearly has
to be justified (see \cite{HSS:zbb,RH:rhein}).  The whole calculation
would not have been possible without the computer program {\tt
  EXP}~\cite{Sei:dipl}. As compared to {\tt lmp}, it also performs the
\hmp\ and its combination with the \lmp. Meanwhile the results of
\cite{HSS:zbb} have been confirmed by a different
approach~\cite{FJTV:zbb}.

{\bf \bld{t\to bW} to \bld{\order{\alpha_s^2}}:} the problem here is
that the contributing three-loop diagrams are actually on-shell,
$q^2=M_t^2$. Two different methods have been used to compute this
process: The first calculation \cite{CzaMel:twb} used asymptotic
expansions in the limit $1-M_b^2/M_t^2\ll 1$, the second one
\cite{CHSS:twb} evaluated the off-shell diagrams for $q^2\ll M_t^2$ with
the help of the \hmp, taking the limit $q^2\to M_t^2$ after performing a
Pad\'e approximation (see \cite{CHSS:twb,CHSS:tampere} for details). The
perfect agreement and the fairly high accuracy of the results in both
approaches is a clear demonstration of the power of asymptotic
expansions.

{\bf \bld{e^+e^-\to t\bar t} near threshold:} the development of
asymptotic expansions in the threshold limit allowed to compute this
process up to ${\cal O}(v^2$, $\alpha_s v$, $\alpha_s^2)$
\cite{eett_thr}, where $v$ is the velocity of the top quarks in the cms
system. It turned out that the corrections are huge and exhibit a strong
dependence on the renormalization scale. The interpretation of this
result and its implications are still an ongoing discussion (see
\cite{jez:ustron} and references therein).

{\bf \bld{\msbar} to pole mass conversion at \bld{\order{\alpha_s^3}}:}
among other things, the progress in threshold calculations mentioned
above makes it very important to be able to express the $\msbar$ mass in
terms of the pole mass at $\order{\alpha_s^3}$. The corresponding
calculation of this relation was based on asymptotic
expansions in the limit of small and large quark mass and interpolated
the result for the actual on-shell diagrams with the help of Pad\'e
approximations \cite{CheSte99}.

{\bf \bld{\mu} decay and semileptonic \bld{b} decays:} for special final
state configurations the decay $b\to cl\bar\nu$ was computed up to
$\order{\alpha_s^2}$ using asymptotic expansions in the limit
$1-M_c^2/M_b^2\ll 1$~\cite{bdec:zerorec}. For $b\to ul\bar\nu$, the full
inclusive result was obtained through a four-loop calculation, where the
actual integrals were calculated by performing an asymptotic expansion
and resumming the full series~\cite{TvR:bdecay}. The same technique was
applied to the 2-loop QED corrections to muon decay \cite{RS:mudecay}.
Using an approach similar to the one of \cite{CHSS:twb} (see above), the
results of \cite{RS:mudecay,TvR:bdecay} were recently
confirmed~\cite{SeiSte:mudecay}.


\section{Back to OPE: RG improvement}\label{sec::rgimp}
The advantages of the approach of Sections~\ref{sec::regions} and
\ref{sec::lmphmp} in comparison to the one of Section~\ref{sec::ope}
when calculating Feynman diagrams are obvious: In the OPE language one
needs to define a full set of operators for every order in $m^2$ and to
apply the appropriate projectors in order to obtain the coefficient
functions. Using the \lmp, all operations can be performed on the level
of Feynman diagrams. The depth of the expansion in $m^2$ is just a
matter of evaluating the Taylor expansions to sufficiently high orders.
However, the loss of contact to the quantum field theoretic level has
its price. In this section two advantages of the OPE language over the
strict application of the \lmp\ will be described.


\subsection{Resummation of logarithms\cite{ref::resum}}\label{sec::resum}
Consider the expansion of the polarization function w.r.t.\ small
$m^2/q^2$ (see Eq.~(\ref{eq::pi0v})). The coefficients
of this series contain logarithms of the form $\ln(\mu^2/m^2)$ and
$\ln(-\mu^2/q^2)$. The former arise from the tadpole diagrams (in the
OPE language: the vacuum expectation values; in the \lmp: the
co-subgraphs), the latter from the massless propagator diagrams
(coefficient functions/hard subgraphs).  Working at fixed order in
perturbation theory, one has to choose some value for $\mu$ in order to
make predictions for physical quantities. It is clear that one should
choose $\mu$ such that the coefficients of the perturbative series are
small. However, with the two types of logarithms above this might be
impossible. In the OPE approach this problem can be solved in the
following way: the operators fulfill the renormalization group (RG)
equation
\begin{equation}
\oddo{}{\mu^2}\opo_n = \sum_n\gamma_{nm}\opo_m\,,
\label{eq::mudmuopo}
\end{equation}
where $\gamma_{nm}$ is their anomalous dimension related to the
renormalization matrix $Z_{nm}$ of (\ref{eq::opoB}) by
\begin{equation}
\gamma_{nm} = \sum_k\left(\oddo{}{\mu^2}Z_{mk}\right)(Z^{-1})_{kn}\,.
\end{equation}
The solution of (\ref{eq::mudmuopo}) is of the form
\begin{equation}
\opo_n(\mu) = \sum_m R_{nm}\opo_m(\mu_0)\,,
\end{equation}
where $R_{nm}$ depends on $\mu$ and $\mu_0$ only
implicitly through its dependence on $\alpha_s$.
One can now rewrite the OPE as
\begin{equation}
\Pi(q^2) = \sum_nC_n(\mu)\opo_n(\mu) =
\sum_{n,m}C_n(\mu)R_{nm}\opo_m(\mu_0)\,.
\end{equation}
Setting $\mu^2=q^2$ and $\mu_0^2=m^2$ removes both types of logarithms
discussed above.  Using the diagram-wise expansions of
Sections~\ref{sec::regions} or \ref{sec::lmphmp}, such kind of
resummation is not possible in a straightforward way.


\subsection{Reconstruction of higher orders in $\alpha_s$}\label{sec::a3m4}
Consider the expansion of $\Pi(q^2)$ in small $m^2/q^2$ at $(l+1)$-loop
level. It appears that the only proper $(l+1)$-loop diagrams that
contribute arise from the naive Taylor expansion, or in other words,
from the trivial operators ${\bf m}^{2n}$.  All the additional terms are
products of diagrams with a lower number of loops.  In the following we
will show that this fact allows to derive the physically relevant
quantity $R = \Im\Pi$ at $(l+1)$-loop level from an $l$-loop
calculation assuming the OPE approach of Section~\ref{sec::ope}.

To be specific, consider the calculation of the $m^4$ terms of
$\Pi(q^2)$. The dimension-4 piece of the OPE fulfills the following RG
equation:
\begin{equation}
\oddo{}{\mu^2}\sum_{n=1}^3C_n\opo_n = 0\,.
\label{eq::rgedim4}
\end{equation}
The effect of the derivative on the $\opo_n$ is again determined by
(\ref{eq::mudmuopo}). 
On the other hand, because the $C_n$ do not depend
on masses, we can write
\begin{equation}
\oddo{}{\mu^2}C_n = \left(\doverd{L} +
  \alpha_s\beta\doverd{\alpha_s}\right)C_n\,,
\end{equation}
where $L=\ln(-\mu^2/q^2)$.
Inserting this into (\ref{eq::rgedim4}), collecting the coefficients of
$\opo_3$ and using $\gamma_{33} = 4\gamma^m$ (this can be seen by
recalling $\oddo{}{\mu^2} m = \gamma^m m$), one obtains \cite{CheKue94}:
\begin{equation}
\doverd{L}C_3 = -4\gamma^m C_3 - \alpha_s\beta\doverd{\alpha_s} C_3 -
\sum_{n=1,2} \gamma_{n3}C_n\,.
\end{equation}
Performing an $l$-loop calculation, $C_1$ and $C_2$ are known to order
$\alpha_s^{l}$, $C_3$ only to $\alpha_s^{l-1}$. But since $\beta$ and
$\gamma^m$ do not contain terms of order $\alpha_s^0$, the r.h.s.\ is
known to $\order{\alpha_s^{l}}$ (assuming that the anomalous dimensions
are known to appropriate order), and so is the l.h.s.  The logarithmic
terms (and thus the desired imaginary part) of $C_3$ can therefore be
obtained to $\order{\alpha_s^l}$ by trivial integration.

This strategy was first followed in \cite{CheKue94} to derive the
$m^4\alpha_s^2$ terms of $R(s)$. The extension to $m^4\alpha_s^3$ was
performed in \cite{CHK:a3m4} (see also \cite{CHK:tampere}).



\section{Conclusions}
In this lecture we have discussed the methods for asymptotic expansions
of Feynman integrals.  The field theoretical approach via operator
product expansion has been compared to the diagram-wise methods, and the
advantages of both strategies have been outlined. We hope that the
importance of the numerous applications --- we could sketch only a few
of them --- has convinced the reader of the power and flexibility of
these expansions.


\section*{Acknowledgments}

I would like to thank K.G.~Chetyrkin, J.H.~K\"uhn, T.~Seidensticker, and
M.~Steinhauser for fruitful collaboration on various topics and careful
reading of the manuscript. I am indebted to A.~Czarnecki, K.~Melnikov,
and A.~R\'etey for several enlightening discussions and useful comments
on the text.  I would further like to thank the organizers of the school
for the invitation and the pleasant atmosphere.  This work was supported
by the {\it Deutsche Forschungsgemeinschaft}.



\def\app#1#2#3{{\it Act.~Phys.~Pol.~}{\bf B #1} (#2) #3}
\def\apa#1#2#3{{\it Act.~Phys.~Austr.~}{\bf#1} (#2) #3}
\def\cmp#1#2#3{{\it Comm.~Math.~Phys.~}{\bf #1} (#2) #3}
\def\cpc#1#2#3{{\it Comp.~Phys.~Commun.~}{\bf #1} (#2) #3}
\def\epjc#1#2#3{{\it Eur.\ Phys.\ J.\ }{\bf C #1} (#2) #3}
\def\fortp#1#2#3{{\it Fortschr.~Phys.~}{\bf#1} (#2) #3}
\def\ijmpc#1#2#3{{\it Int.~J.~Mod.~Phys.~}{\bf C #1} (#2) #3}
\def\ijmpa#1#2#3{{\it Int.~J.~Mod.~Phys.~}{\bf A #1} (#2) #3}
\def\jcp#1#2#3{{\it J.~Comp.~Phys.~}{\bf #1} (#2) #3}
\def\jetp#1#2#3{{\it JETP~Lett.~}{\bf #1} (#2) #3}
\def\mpl#1#2#3{{\it Mod.~Phys.~Lett.~}{\bf A #1} (#2) #3}
\def\nima#1#2#3{{\it Nucl.~Inst.~Meth.~}{\bf A #1} (#2) #3}
\def\npb#1#2#3{{\it Nucl.~Phys.~}{\bf B #1} (#2) #3}
\def\nca#1#2#3{{\it Nuovo~Cim.~}{\bf #1A} (#2) #3}
\def\plb#1#2#3{{\it Phys.~Lett.~}{\bf B #1} (#2) #3}
\def\prc#1#2#3{{\it Phys.~Reports }{\bf #1} (#2) #3}
\def\prd#1#2#3{{\it Phys.~Rev.~}{\bf D #1} (#2) #3}
\def\pR#1#2#3{{\it Phys.~Rev.~}{\bf #1} (#2) #3}
\def\prl#1#2#3{{\it Phys.~Rev.~Lett.~}{\bf #1} (#2) #3}
\def\pr#1#2#3{{\it Phys.~Reports }{\bf #1} (#2) #3}
\def\ptp#1#2#3{{\it Prog.~Theor.~Phys.~}{\bf #1} (#2) #3}
\def\ppnp#1#2#3{{\it Prog.~Part.~Nucl.~Phys.~}{\bf #1} (#2) #3}
\def\sovnp#1#2#3{{\it Sov.~J.~Nucl.~Phys.~}{\bf #1} (#2) #3}
\def\tmf#1#2#3{{\it Teor.~Mat.~Fiz.~}{\bf #1} (#2) #3}
\def\tmp#1#2#3{{\it Theor.~Math.~Phys.~}{\bf #1} (#2) #3}
\def\yadfiz#1#2#3{{\it Yad.~Fiz.~}{\bf #1} (#2) #3}
\def\zpc#1#2#3{{\it Z.~Phys.~}{\bf C #1} (#2) #3}
\def\ibid#1#2#3{{ibid.~}{\bf #1} (#2) #3}

\end{document}